\documentclass[aps,reprint,showpacs,preprintnumbers,amsmath,amssymb,prb]{revtex4-1} 

\usepackage{graphicx}
\graphicspath{{figs/}{./}}

\usepackage{color}
\newcommand\FigureFile[1] {#1.eps}
\DeclareGraphicsExtensions{pdf}

\usepackage{dcolumn}
\usepackage{bm}
\usepackage{color}
\usepackage{multirow}

\newcommand\eq[1]                              
{
\begin{align*}
#1
\end{align*}
}

\newcommand\eql[2] 
{
\begin{equation}\label{#1}
\begin{split}
#2
\end{split}
\end{equation}
}

\newcommand\eqsl[1]                            
{
\begin{align}
#1
\end{align}
}

\newcommand\eqssl[2]                      
{
\begin{subequations}\label{#1}
\begin{align}
#2
\end{align}
\end{subequations}
}

\newcommand\Eq[1]      {Eq.~\eqref{#1}}

\definecolor{xmgrace-green4}{rgb}{0.0,0.55,0.0}
\definecolor{Green}{rgb}{0.2,0.96,0.2}
\definecolor{Remarks}{rgb}{1,0.3,0.3}
\definecolor{Extra}{rgb}{0.2,0.2,1}
\definecolor{Blue}{rgb}{0.2,0.3,1}
\definecolor{Black}{rgb}{0,0,0}

\newcommand\compPackage[1] {{\footnotesize{#1}}}
\newcommand\NWCHEM     {\compPackage{NWCHEM}}
\newcommand\QUANTUMESPRESSO   {\compPackage{QUANTUM ESPRESSO}}

\newcommand\PWSCF      {\compPackage{PWSCF}}

\newcommand\COMMENTED[1] {}

\begin{document}

\title{\emph{Ab Initio} Many-body Study of Cobalt Adatoms Adsorbed on Graphene}

\author{Yudistira Virgus}
\email{yvirgus@email.wm.edu}
\affiliation{Department of Physics, College of William and Mary,
Williamsburg, Virginia 23187-8795, USA}

\author{Wirawan Purwanto}
\affiliation{Department of Physics, College of William and Mary,
Williamsburg, Virginia 23187-8795, USA}

\author{Henry Krakauer}
\affiliation{Department of Physics, College of William and Mary,
Williamsburg, Virginia 23187-8795, USA}

\author{Shiwei Zhang}
\affiliation{Department of Physics, College of William and Mary,
Williamsburg, Virginia 23187-8795, USA}

\date{\today}

\begin{abstract}

Many recent calculations have been performed to study 
a Co atom adsorbed on graphene, with significantly varying results on
the nature of the bonding.
We use auxiliary-field quantum Monte Carlo (AFQMC) and a size-correction embedding scheme to accurately calculate the binding energy of Co on graphene.
We find that as a function of the distance $h$ between the Co atom and the six-fold hollow site,
there are three distinct ground states corresponding to three electronic configurations of the Co atom. Two of these states provide binding and exhibit a double-well feature with nearly equal binding energy of $0.4$~eV at $h = 1.51$ and $h = 1.65$~\AA, corresponding to low-spin $^2$Co ($3d^{9}4s^{0}$) and high-spin $^4$Co ($3d^{8}4s^{1}$), respectively.

\end{abstract}


\maketitle

Since its discovery, graphene has been the subject of intense efforts to adapt it for a variety of promising applications due to its unique and exceptional intrinsic properties. \cite{Geim2007, Castro2009}
One potential application is for use in spintronic devices. \cite{Zhang_spin_2011, Pesin_2012, Kotov2012}
However, external methods are required to induce magnetism on graphene, since pristine graphene is nonmagnetic.
One proposal is to adsorb transition metal atoms to provide localized magnetic moments in graphene.
Single Co atoms on graphene have been extensively studied recently, 
 \cite{Yagi2004, Mao2008, Johll2009, Wehling2010-a, Wehling2010-b, Jacob2010, Cao2010, Valencia2010, Chan2011, Liu2011, Sargolzaei2011, Ding2011, Wehling2011,Rudenko2012}
 and possible Kondo effects have been considered. \cite{Mattos2009,Brar2011} 
The study of Co/graphene is thus of great interest both from a fundamental and applied perspective.
 
Theoretical treatments of Co/graphene systems have largely been done at the density functional theory (DFT) level with local or semi-local functionals, or with an empirical Hubbard on-site repulsion $U$ (DFT$+U$). 
\cite{Yagi2004, Mao2008, Johll2009, Wehling2010-a, Wehling2010-b, Jacob2010, Cao2010, Valencia2010, Chan2011, Liu2011, Sargolzaei2011, Ding2011, Wehling2011} 
However, the applicability of methods based on independent-electron approximations in such systems is unclear, since electron correlation effects can be significant. 
Indeed, widely varying results have been reported for the nature of the magnetic state and binding of Co as a function of adsorption height. 
DFT calculations with the generalized gradient approximation (GGA)  \cite{Perdew1996} predict \cite{Yagi2004, Mao2008, Johll2009, Wehling2010-a, Wehling2010-b, Jacob2010, Cao2010, Valencia2010, Chan2011, Liu2011, Sargolzaei2011, Ding2011, Wehling2011} 
an equilibrium height of $h_{\textrm{eq}} \sim 1.5$ \AA\ above the six-fold hollow site, with a low-spin Co atom configuration ($S = 1/2$). 
A different functional, the hybrid B3LYP, \cite{Becke1993} predicts \cite{Jacob2010} an equilibrium height of $h_{\textrm{eq}} \sim 1.9$ \AA\ at the hollow site, with a high-spin configuration ($S = 3/2$). 
Results from the GGA$+U$ approach have shown sensitivity to the choice of the parameter $U$ which leads to different spin configuration, equilibrium height, and equilibrium site for different values of $U$. \cite{Wehling2010-b, Chan2011, Wehling2011}
A recent quantum chemistry calculation using the 
 complete active space self-consistent field method gives a state from the van der Waals (vdW) interaction (high-spin  $3d^{7}4s^{2}$ state)  as the global minimum, with $h_{\textrm{eq}} \sim 3.1$ \AA.\  \cite{Rudenko2012}
These contrasting results strongly indicate the need for a more accurate \emph{ ab initio} treatment of electron correlations in Co/graphene.

In this paper,  we use the auxiliary-field quantum Monte Carlo (AFQMC) method \cite{Zhang1997_CPMC, Zhang2003} to investigate the binding energy and electronic properties of 
Co/graphene.
We focus on the hollow site which is the most favorable adsorption site according to most DFT calculations. 
Contrary to prior calculations, we find that as the Co atom approaches the graphene sheet, it experiences two magnetic transitions which lead to three distinct ground-state electronic configurations. 
One of these configurations corresponds to the vdW interaction. 
The other two configurations arise from a strong orbital hybridization and provide binding with a double-well feature.

\begin{figure}[tbp]
\includegraphics[scale=0.93]{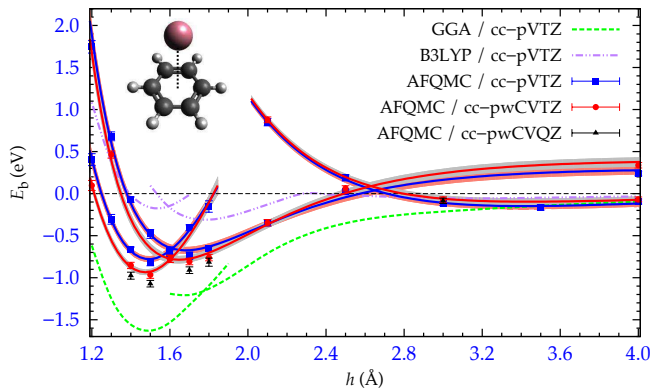}
\caption{\label{fig:Co-benzene-AFQMC-GGA-B3LYP}
(Color online) Binding energy of Co on C$_6$H$_6$ as a function of Co adsorption height $h$ at the six-fold site for different methods. 
For AFQMC, left, middle, and right curves correspond to nominal $3d^{9}4s^{0}$, $3d^{8}4s^{1}$, and $3d^{7}4s^{2}$ Co configurations, respectively. AFQMC results include Trotter time step extrapolation.
For DFT results, the left and right curves correspond to $3d^{9}4s^{0}$ and $3d^{8}4s^{1}$ Co configurations, respectively. 
The shaded area on the AFQMC Morse fits reflects one standard deviation statistical errors.
}
\end{figure}

Since strong electron-electron interactions are expected to be spatially localized in the immediate vicinity of the Co atom, 
we use a size-correction embedding scheme (ONIOM 
\cite{Sven1996}) to 
accelerate convergence and reach large system sizes in the many-body calculations.
In this approach, the ``near" region in the vicinity of the Co atom is modeled by a relatively small number of atoms, using a highly accurate many-body method like AFQMC, while size corrections are treated using a lower level of theory like DFT.
For the near region, we chose the Co atom and its six nearest neighbor substrate C atoms, with their dangling bonds terminated by H
atoms, resulting in a Co/C$_{6}$H$_{6}$ benzene-like system (see the inset in Fig.~\ref{fig:Co-benzene-AFQMC-GGA-B3LYP}).
The size-corrected binding energy of the Co/graphene system is then given by
\eql{eq:ONIOM}
{
       E_{\textrm{\scriptsize{{b}, ONIOM}}} = E_{\textrm{\scriptsize{{b}, AFQMC}}}^{\textrm{\scriptsize{Co/C$_6$H$_6$}}} + ( E_{\textrm{\scriptsize{{b}, DFT}}}^{\textrm{\scriptsize{Co/graphene}}} -  E_{\textrm{\scriptsize{{b}, DFT}}}^{\textrm{\scriptsize{Co/C$_6$H$_6$}}})       
       \,,
}
which we will calculate as a function of $h$, the perpendicular distance between Co atom and the substrate, for each spin multiplicity of the Co atom. 
For each substrate, $E_{\textrm{\scriptsize{{b}}}}$ is defined as $E_{\textrm{\scriptsize{{b}}}} \equiv E^{\textrm{\scriptsize{Co/substrate}}} -  E^{\textrm{\scriptsize{Co}}} - E^{\textrm{\scriptsize{substrate}}} $.
The Co/C$_6$H$_6$ C--C bond length was fixed to that of graphene, $1.42$ \AA, which is 
only slightly larger than the experimental benzene value of $1.40$ \AA, while the distance to the H ``link atom," the C--H bond length, was set to $1.09$ \AA,\  which is the predicted geometry by GGA for the corresponding C--C bond length.
Previous studies have shown little sensitivity to the link-atom bond distance. \cite{Caricato2010}
Our AFQMC calculations were all done for fixed substrate geometries. We will consider the effect of 
substrate geometry relaxation with the assistance of DFT calculations, as discussed below.

The AFQMC method  \cite{Zhang1997_CPMC, Zhang2003} evaluates the ground state properties of a many-body Hamiltonian stochastically, using random walks with Slater determinants expressed in a chosen one-particle basis. 
Although AFQMC is an exact method in principle, the fermion sign problem causes an exponential growth of the Monte Carlo variance.
The problem is controlled using a constraint on the overall phase of the Slater determinants
during the random walks, the
phaseless approximation, \cite{Zhang2003} that relies on a trial wave function. 
In extensive benchmarks in both strongly correlated lattice models and  
molecular and crystalline systems, the method 
has shown excellent agreement with exact and/or experimental results. \cite{Zhang1997_CPMC, Zhang2003,AlSaidi2006a,AlSaidi2006b,AlSaidi2006c,AlSaidi2007b,Purwanto2008}
This is consistent with expectations from analysis of the origin of the sign problem and the nature of the constraint. \cite{Zhang1997_CPMC, Zhang2003}
In most calculations to date on realistic systems (molecules and solids), trial wave functions of a single Slater determinant from Hartree-Fock 
or DFT have been used and have been shown to give results whose accuracy 
is comparable to the best many-body methods, for example 
coupled-cluster CCSD(T) in molecules. 
In this paper, we use the phaseless AFQMC method working with standard Gaussian single-particle basis sets, \cite{AlSaidi2006b} and a recent
implementation of the frozen-core approximation to treat the inner core electrons. \cite{Purwanto-FC}  

We first report AFQMC results for Co on a C$_6$H$_6$ substrate.
In themselves, these results provide a direct and systematic benchmark of other computational methods.
The binding energy curves of Co/C$_{6}$H$_{6}$ from AFQMC and DFT (GGA and B3LYP) are shown in Fig.~\ref{fig:Co-benzene-AFQMC-GGA-B3LYP}.
AFQMC results show that, with decreasing $h$, the ground-state electronic configuration of the Co atom undergoes two transitions resulting in three different configurations: high-spin $3d^{7}4s^{2}$,  high-spin $3d^{8}4s^{1}$, and low-spin $3d^{9}4s^{0}$ states, respectively.
Only two DFT ground-state configurations are found,
a high-spin $3d^{8}4s^{1}$ for large $h$ and a low-spin $3d^{9}4s^{0}$ for small $h$. 
This is because both DFT functionals incorrectly predict  $3d^{8}4s^{1}$ to be the ground state configuration for the free Co atom.
Both GGA and B3LYP predict low- and high-spin relative minima in the vicinity of the AFQMC predictions, but GGA severely overestimates the well-depths, which are underestimated by B3LYP. 

The AFQMC calculations were done with our recently implemented  frozen-core approximation,  
\cite{Purwanto-FC} thus avoiding the need for pseudopotentials; only the most tightly bound core states
were frozen:  Co($1s,2s,2p$) and C($1s$).
 The potential energy curves (PECs) are obtained by AFQMC calculations with fixed $S_z$, 
 in which the numbers of electrons with $\uparrow$- and $\downarrow$-spins are preset.
 The Co spin configuration in the different PECs is identified by that of the trial wave function
  $\Psi_\mathrm{T}$. \cite{Zhang2003}
 Thus these are nominal states and do not imply literal spin configuration of Co in the many-body 
 ground state. 
Typical AFQMC runs used $\simeq 5000$~walkers and a Trotter time step $\Delta \tau = 0.01~\mathrm{Ha}^{-1}$, and final results were extrapolated to the $\Delta \tau \rightarrow 0$ limit.
All AFQMC calculations for Co/C$_{6}$H$_{6}$ used single-determinant unrestricted Hartree-Fock (UHF)  $\Psi_\mathrm{T}$.
Previous experience indicates that such AFQMC calculations are very accurate, 
\cite{AlSaidi2006b,AlSaidi2006c,AlSaidi2007b,Purwanto2008,Purwanto2011} including for systems containing transition metal atoms. \cite{AlSaidi2006a}  Future study using multi-determinant
 $\Psi_\mathrm{T}$ is warranted, however, given the new territory being explored here with QMC.
DFT and HF calculations that use Gaussian basis sets are performed using {\NWCHEM}. \cite{Valiev2010}

\begin{table}[t]
\caption{\label{tbl:Co-benzene}
Calculated binding energies $E_{\textrm{b}}$ and adsorption heights $h$ of Co on C$_6$H$_6$ at the three local minima shown in Fig.~\ref{fig:Co-benzene-AFQMC-GGA-B3LYP} (distances in
\AA\ and energies in eV).
Tabulated $E_{\textrm{b}}$ at the low-spin $h = 1.47$~\AA\ and high-spin  $h=1.65$~\AA\ minima are CBS extrapolations. $E_{\textrm{b}}$ at the van der Walls (vdW) 
$h=3.4$~\AA\ minimum is essentially converged at the QZ level. 
}
\begin{ruledtabular}
\begin{tabular}{c c c c c c c}
\vspace{-0.8em}
\\
                       & \multicolumn{2}{c}{AFQMC (CBS)}
                        & \multicolumn{2}{c}{GGA} 
                        & \multicolumn{2}{c}{B3LYP} \\ \cline{2-3} \cline{4-5} \cline{6-7}
\vspace{-0.8em}
\\
                        & $h_{\textrm{eq}}$ & $E_{\textrm{b}}$ & $h_{\textrm{eq}}$ & $E_{\textrm{b}}$ & $h_{\textrm{eq}}$  & $E_{\textrm{b}}$ \\

\hline
\vspace{-0.8em}
\\
$S=1/2$
                    & $1.47$    & $ -1.07(6)$ &  $1.49$ & $-1.63$   & $1.54$  & $-0.17$              \\
\hline
\vspace{-0.8em}
\\
$S=3/2$ \\

$3d^{8}4s^{1}$              
                    &  $1.65$     &  $-0.92(5)$   &  $1.66$ & $-1.21$  &   $1.78$ & $-0.31$\\
vdW
                    &   $3.4$   &  $-0.10(3)$   &  -- & -- & -- & -- \\
\end{tabular}
\end{ruledtabular}
\end{table}

Care was taken to remove finite basis set error in the many-body results. 
The following basis sets were used in AFQMC calculations for most $h$. 
The Co atom used the correlation-consistent core-valence cc-pwCVTZ basis set, where "core" refers to the Co $3s,3p$ semicore states. 
For C and H atoms, valence-only cc-pVTZ and cc-pVDZ were used, respectively.
For several geometries near the minima,  the Co(cc-pwCVQZ) basis set was used to obtain extrapolation to the complete basis set (CBS) limit.
Not surprisingly, while the DFT results are converged by the Co(cc-pVTZ) level,
AFQMC is not yet fully converged even at the Co(cc-pwCVQZ) level.
To estimate the effect of the CBS extrapolation,  we used the procedure in Ref. \onlinecite{Helgaker1997}: an exponential form for the HF contribution to the total energy 
and an inverse-third-power form for the correlation energy. 
Extrapolation to the CBS limit lowers the binding energy near the minima by  $0.13$\,eV from the TZ result and $0.03$\,eV
 from that of the QZ basis.
Trotter time step extrapolations were obtained from results for a smaller basis set [cc-pVTZ for Co and cc-pVDZ for C and H] and applied to the $E_{\textrm{b}}$ results for the larger basis sets.
Final, fully extrapolated AFQMC results at the three minima are tabulated in Table~\ref{tbl:Co-benzene}.
The global minimum in AFQMC is the low-spin $3d^{9}4s^{0}$ state with binding energy \mbox{$E_\mathrm{b} = -1.07(6)$\,eV}, as seen in the Table. 
The high-spin minimum has only a slightly smaller \mbox{$E_\mathrm{b}=-0.92(5)$\,eV}.
In the vdW region, the system is barely bound with \mbox{$E_\mathrm{b}=-0.10(3)$\,eV}.

\begin{figure}[tbp]
\includegraphics[scale=0.93]{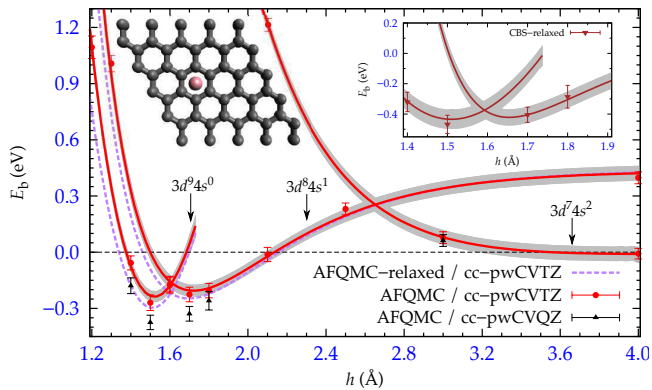}
\caption{\label{fig:Co-benzene-graphene-oniom-CBS}
(Color online) Binding energy of Co atom on graphene as a function of $h$. 
Left, middle, and right curves correspond to $3d^{9}4s^{0}$, $3d^{8}4s^{1}$, and $3d^{7}4s^{2}$ Co configurations, respectively. 
Shaded areas are one-$\sigma$ statistical error bars.
The left inset shows the structure of Co on graphene in 5 $\times$ 5 supercell. The right inset shows the binding energy after CBS extrapolation and substrate relaxation (see text). The shaded areas in the right inset include both statistical and systematic errors.
}
\end{figure}

Results are then obtained for Co/graphene using the ONIOM embedding scheme.
The finite-size correction [second term in \Eq{eq:ONIOM}] is applied to the AFQMC $E_\mathrm{b}$ curve in Fig.~\ref{fig:Co-benzene-AFQMC-GGA-B3LYP}. The results 
 are shown in Fig.~\ref{fig:Co-benzene-graphene-oniom-CBS}.
To obtain $E_{\textrm{\scriptsize{{b}, DFT}}}^{\textrm{\scriptsize{Co/graphene}}}$, we used DFT-GGA as implemented in the {\PWSCF} code of the {\QUANTUMESPRESSO} package,\cite{Giannozzi2009} with periodic boundary conditions and ultrasoft pseudopotentials. \footnote{We used the pseudopotentials C.pbe-rrkjus.UPF and Co.pbe-nd-rrkjus.UPF from http://www.quantum-espresso.org}
A  \mbox{5 $\times$ 5} in-plane supercell was used, which contains 50 C atoms and a Co atom; the in-plane lattice parameter was $12.3$ \AA,\  while the periodic repeat distance perpendicular to the graphene plane was set to $15$ \AA.
A planewave basis kinetic energy cutoff  of $E_\mathrm{cut}=45$~Ry and 
a charge density cutoff $360$~Ry were used for all geometries. 
Brillouin-zone sampling used a  $\Gamma$-centered  \mbox{$ 4 \times 4 \times 1$}  $k$-point grid and a Gaussian smearing width of $0.04$ eV.
The ONIOM $E_\mathrm{b}$ correction was obtained from similar {\PWSCF} calculations for the clean \mbox{$ 5 \times 5 $}  graphene supercell;
the energy of an isolated Co atom was obtained using a large supercell with single $k$-point sampling. 
Approximate relativistic corrections are included in our results via ONIOM as 
the GGA calculations are scalar-relativistic, although the correction is not perfect due to the absence of the vdW curve in GGA.
The lines in Fig.~\ref{fig:Co-benzene-graphene-oniom-CBS} are Morse fits to the cc-pwCVTZ AFQMC results. 

\begin{figure}[tbp]
\includegraphics[scale=0.93]{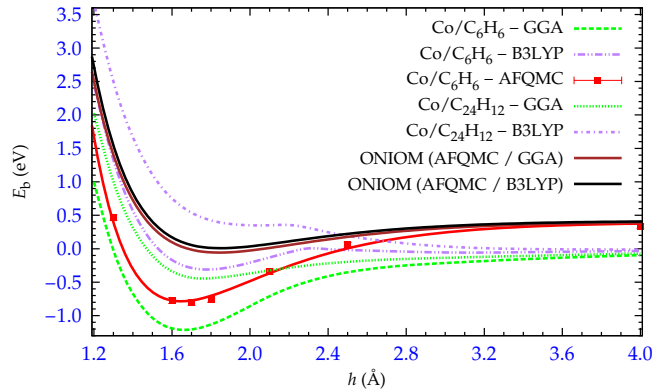}
\caption{\label{fig:Co-benzene-coronene-oniom}
(Color online) ONIOM size corrections and the binding energies of Co/C$_{6}$H$_{6}$ and Co/C$_{24}$H$_{12}$ systems in  the $3d^{8}4s^{1}$ state. The two ONIOM curves are basically identical and show insensitivity to the choice of DFT flavors. The corrections are applied to the $3d^{8}4s^{1}$ AFQMC/cc-pwCVTZ 
binding energy curve in Fig.~\ref{fig:Co-benzene-AFQMC-GGA-B3LYP}. Similar independence on DFT functional is found for the other spin states.
 }
\end{figure}

It is reassuring to note that 
the size correction in \Eq{eq:ONIOM} is essentially independent of the choice of DFT exchange-correlation functional.
This is illustrated, for GGA and B3LYP, in Fig.~\ref{fig:Co-benzene-coronene-oniom}, using a coronene-like C$_{24}$H$_{12}$ substrate which is comprised of six joined C$_{6}$ rings with outer H terminations.
(B3LYP calculations for the $5\times5$ supercell were time consuming and difficult to converge.) 
As Fig.~\ref{fig:Co-benzene-coronene-oniom} illustrates, while GGA and B3LYP 
show large differences between their $E_{\textrm{b}}$ curves, the size correction in \Eq{eq:ONIOM} is essentially independent of which is used.

We examined substrate relaxation effect by comparing the relaxed and unrelaxed \mbox{5 $\times$ 5} {\PWSCF} supercell results and including it as  an additional ONIOM ``layer."
For this purpose,  the six C atoms nearest Co in the relaxed substrates were allowed to relax only in the in-plane direction. 
The value of $h$ was defined in relation to these atoms; the remainder of the C atoms were completely relaxed in $C_{2v}$ symmetry. 
Relaxation was considered complete when the force on all atoms, except the restricted atoms, was less than 0.02 eV/\AA.
Near the double well minima, fully relaxing  all the atoms had little additional effect near the low-spin (high-spin) minimum: in- and out-of-plane distortions are  
$< 0.015$ $(0.011)$ \AA \ and $< 0.01$ $(0.002)$ \AA, respectively.
Substrate relaxation lowers the binding energy by about $0.05$\,eV near the minima.

The inset of Fig.~\ref{fig:Co-benzene-graphene-oniom-CBS} shows the binding energy curves near the double well feature, after CBS extrapolation
and substrate relaxation effects have been included.
The two wells in the Co/graphene PEC's 
have comparable binding energies of 
$-0.4$\,eV. 
The vdW region shows no binding within AFQMC statistical resolution.
STM experiments could, in principle, detect the spin-state of Co atoms on graphene.  \cite{Brune2009}
Recently, controllable ionization and screening of Co atoms on graphene via scanning tunneling microscopy (STM) have been observed. \cite{Brar2011}
Kondo screening is generally considered for effectively  $S=1/2$ impurity systems.
Higher spin states can be observed, however, in the presence of magnetic anisotropy, 
if it results in a low-lying degenerate doublet ground state, 
as was observed for individual Co atoms adsorbed on 
Cu(100) crystals that are covered by a monolayer of copper nitride (Cu$_2$N). \cite{Otte2008}
Mattos \cite{Mattos2009} reported STM observations of Kondo signatures for Co/graphene.
Brar {\em et al.},\cite{Brar2011} however,  measured a  Kondo-like dip
feature (with a 5 meV half-width in $dI/dV$) for Co on back-gated graphene/SiO$_2$, which they instead attributed to vibrational inelastic tunneling.
To model this they performed DFT supercell calculations for free standing hollow-site Co/\mbox{$(4 \times 4$)-graphene} and found in-plane vibrational modes of 12 and 27 meV, and out-of-plane modes of 17, 40 and 53 meV,  \cite{Brar2011}
the lowest of which are roughly commensurate with the observed 5 meV width.
Within the statistical resolution of the AFQMC double wells in the inset of 
Fig.~\ref{fig:Co-benzene-graphene-oniom-CBS}, 
both low- and high-spin minima have the  same curvature, 
corresponding to an out-of-plane frequency range 16~--~58 meV, qualitatively similar to the DFT frequencies.
At liquid He temperatures where the STM experiments are performed, tunneling between the minima in Fig.~\ref{fig:Co-benzene-graphene-oniom-CBS}  can be neglected, based on a barrier height of $0.04$\ eV.
Experimental determinations are further complicated, however, by indications that the charge state of single Co atoms on graphene switches in proximity
to the STM tip. \cite{Wang2012}
To the best of our knowledge, current experiments for Co/graphene have not yet determined the spin state of individual Co atoms adsorbed on graphene.  
Our results are consistent with roughly equal occurrence of Co $S=1/2$  and $S=3/2$ atoms populating the two minima, respectively.

In summary, we have presented an \emph{ab initio} many-body study of Co on graphene to address the effect of electron correlations.
We use the AFQMC method with single-determinant trial wave functions to calculate the binding energy curve of Co/C$_{6}$H$_{6}$.
The Co/graphene binding energy was calculated using an ONIOM size-correction procedure.
The size-correction method shows insensitivity to the choice of DFT flavors which suggests that Co/C$_{6}$H$_{6}$ cluster captures most of the correlation effect.
The resulting binding energy curve of Co on graphene exhibits binding with a double-well structure.
Both minima show nearly equal binding energy of $ -0.4$ eV.
The inner well corresponds to a low-spin $S=1/2$ state with a $3d^{9}4s^{0}$ electronic configuration for Co atom, while the outer well is characterized by a high-spin ($S=3/2$) $3d^{8}4s^{1}$  state.
Our results show that the Co/graphene system requires an accurate and careful treatment of
many-body correlation effects. 
Better resolution of the energetics and the characteristics of the ground states will require further work, but the results  suggest a plausible framework which is consistent with recent experimental observations. 
We hope this result will encourage further theoretical and experimental studies of the spin states and Kondo effect in Co on graphene.

This work was supported by 
DOE (DE-FG02-09ER16046), 
NSF (DMR-1006217),
and 
ONR (N000140811235; N000141211042).
An award of computer time was provided by the Innovative and Novel Computational Impact on Theory and Experiment (INCITE) program, using resources of the Oak Ridge Leadership Computing Facility at the Oak Ridge National Laboratory, which is supported by the Office of Science of the U.S. Department of Energy under Contract No. DE-AC05-00OR22725.
We also acknowledge the computing support from the Center for Piezoelectrics by Design.
We would like to thank Enrico Rossi, Fengjie Ma, and Eric Walter for many fruitful discussions. 

\bibliography{AFQMC-bib-entries,Co-graphene,Miscellaneous}

\end{document}